# User Environment Detection with Acoustic Sensors Embedded on Mobile Devices for the Recognition of Activities of Daily Living

Ivan Miguel Pires[1,2,3], Nuno M. Garcia[1,3,4], Nuno Pombo[1,3,4], and Francisco Flórez-Revuelta[5]

[1]Instituto de Telecomunicações, Universidade da Beira Interior, Covilhã, Portugal
[2]Altranportugal, Lisbon, Portugal
[3]ALLab - Assisted Living Computing and Telecommunications Laboratory,
Computer Science Department, Universidade da Beira Interior, Covilhã, Portugal
[4]Universidade Lusófona de Humanidades e Tecnologias, Lisbon, Portugal
[5]Department of Computer Technology, Universidad de Alicante, Spain

impires@it.ubi.pt, ngarcia@di.ubi.pt, ngpombo@di.ubi.pt, francisco.florez@ua.es

**Abstract**

The detection of the environment where user is located, is of extreme use for the identification of Activities of Daily Living (ADL). ADL can be identified by use of the sensors available in many off-the-shelf mobile devices, including magnetic and motion, and the environment can be also identified using acoustic sensors. The study presented in this paper is divided in two parts: firstly, we discuss the recognition of the environment using acoustic sensors (*i.e.,* microphone), and secondly, we fuse this information with motion and magnetic sensors (*i.e.,* motion and magnetic sensors) for the recognition of standing activities of daily living. The recognition of the environments and the ADL are performed using pattern recognition techniques, in order to develop a system that includes data acquisition, data processing, data fusion, and artificial intelligence methods. The artificial intelligence methods explored in this study are composed by different types of Artificial Neural Networks (ANN), comparing the different types of ANN and selecting the best methods to implement in the different stages of the system developed. Conclusions point to the use of Deep Neural Networks (DNN) with normalized data for the identification of ADL with 85.89% of accuracy, the use of Feedforward neural networks with non-normalized data for the identification of the environments with 86.50% of accuracy, and the use of DNN with normalized data for the identification of standing activities with 100% of accuracy.

**Keywords:** Activities of Daily Living (ADL); sensors; mobile devices; accelerometer; gyroscope; magnetometer; microphone; data acquisition; data processing; data cleaning; data fusion; feature extraction; pattern recognition; machine learning.

## 1. Introduction

The acquisition of data related to the Activities of Daily Living (ADL) [1] may be performed with the sensors available in off-the-shelf mobile devices, *e.g.*, the accelerometer, the gyroscope, the magnetometer, the microphone, and the Global Positioning System (GPS) receiver. The acquired data from the sensors available in off-the-shelf mobile devices are related to the movement performed during the activities and the environment where the activities are performed [2] in order to develop a method for the automatic recognition of the ADL as a part of the development of a personal digital life coach [3].

This study proposes the use of the microphone for the recognition of the environment, which is fused with the data acquired from the accelerometer, gyroscope and magnetometer sensors for the recognition of the activities with movement. In continuation of the previous study, available at [4], the main goal of the fusion of the environment recognized with the other sensors' data is to increase the number of ADL recognized using data fusion and artificial intelligence techniques. This study proposes the recognition of ADL, including running, walking, walking on stairs, standing, and sleeping, and the recognition of environments, including bar, classroom, gym, kitchen, library, street, hall, watching TV and bedroom.

These methods are included in the development of a framework for the recognition of ADL and their environments, proposed in [5-7], composed by several modules, such as data acquisition, data processing, data fusion, and artificial intelligence methods. However, the data processing is composed by some steps, such as data cleaning and feature extraction, and the data fusion and artificial intelligence techniques are applied at the same time for the achievement of the final purpose of the recognition of ADL and their environments. The advantages of recognition of the environments are not limited to the increasing of the number of ADL recognized, but it allows the framework to combine the environments with the ADL recognition returning different results, *e.g.,* the user is walking on the street.

The topic related to the recognition of the ADL has some studies available in the literature [8-13], but there are no studies that uses all sensors available on the off-the-shelf mobile devices, however the Artificial Neural Networks (ANN) is one of the most used methods in this topic. Based on our previous studies using motion and magnetic sensors for the development of the framework for the recognition of ADL and their environments [4, 14], this study proposes the creation of several methods to adapt the framework to the number of sensors available in off-the-shelf mobile devices. Some methods using different combinations of sensors are presented in previous studies [4, 14], such as the method using accelerometer data, the method using accelerometer and magnetometer data, and the method using accelerometer, magnetometer and gyroscope sensors. Thus, this study proposes the creation of the method using acoustic data for the recognition of the environments, as well as, different methods fusing the environment recognized with other data sources, such as the method using accelerometer and environment, the method using accelerometer, magnetometer and environment, and the method using accelerometer, magnetometer, gyroscope and environment. For the implementation and testing of these methods, we proposed the use of ANN, exploring the use of three types of ANN, such as the Multilayer Perception (MLP) with Backpropagation implemented with Neuroph [15], the Feedforward neural network with Backpropagation implemented with Encog [16], and Deep Learning implemented with DeepLearning4j [17]. The acquisition of data was performed by people aged between 16 and 60 years old with different lifestyles and the mobile device, correctly positioned on the pocket, for the creation of the datasets composed by the sensors' data. This research included the definition of the correct set of features needed and the best ANN method for the recognition of ADL and environments, verifying that, for the recognition of environments, the best results are achieved with Feedforward neural network with Backpropagation, and, for the recognition of ADL, the best results are achieved with Deep Learning techniques.

The introduction section is concluded in this paragraph, and the remaining sections are organized as follows: Section 2 presents the literature review focused on the use of acoustic sensors for the recognition of ADL and their environments. The methods used for the development of the methods for the implementation in the framework for the recognition of ADL and their environments are presented in the section 3. Section 4 presented the results obtained with the implementation of the different methods. Finally, the discussion about the results and implementation in the framework is presenting in the section 5, presenting the conclusions in the section 6.

## 2. Related Work

There are no studies related to the use of the fusion of the data acquired from all sensors available on off-the-shelf mobile devices, including accelerometer, gyroscope, magnetometer, and microphone, for the recognition of Activities of Daily Living (ADL) and their environments [1], but there are few studies using subsets of these sensors.

The authors of [18] used the Global Positioning System (GPS) receiver, accelerometer, and microphone sensors for the recognition of sleeping, walking, standing, running, and social interaction activities, applying linear and logistic regression methods with several features, including the mean and variance of the accelerometer data and the spectral roll-off of the acoustic data, reporting an accuracy around 90%.

In [19], the authors extracted the minimum, difference between axis, mean, standard deviation, variance, correlation between axis, sum of coefficients, spectral energy and spectral entropy from the accelerometer sensor, and the zero-crossing rate, total spectrum power, sub-band powers, spectral centroid, spectral spread, spectral flux, spectral roll-off, and Mel-Frequency Cepstral Coefficients (MFCC) from the microphone, applied to Support Vector Machine (SVM) and Gradient Boosting Decision Tree methods, in order to recognize sitting on chair, lying, standing, walking, going upstairs, going downstairs, jogging, running, and drinking activities, obtaining results with a reported accuracy between 89.12% and 91.5%.

The authors of [20] recognized several activities, including cycling, cleaning table, shopping, travelling by car, going to toilet, cooking, watching television, eating, driving, working on a computer, reading, and sleeping, using data acquired from the microphone and accelerometer sensors and applying the Gaussian mixture model (GMM) with log power, and MFCC as features, reporting an accuracy of 77.9%.

In [21], the accelerometer and microphone sensors are also used for the recognition of shopping, waiting in a queue, driving, travelling by car, cleaning with a vacuum cleaner, cooking, washing dishes, working at a computer, sleeping, watching television, being a bar, sitting, walking, standing, lying, and standing activities, using J48 decision tree, FT decision tree, LMT decision tree, and IBk lazy algorithm with mean, standard deviation, range, angular degree, and MFCC as features. The reported accuracies are around 90%, where LMT decision tree reports 90.4126%, J48 decision tree reports 90.6553%, IBk lazy algorithm reports 90.7767% and FT decision tree reports 90.6553% [21].

The remaining studies available in the literature using acoustic sensors do not use data fusion techniques, because they only use the microphone signal. Based on the acoustic signal acquired from the microphone, the authors of [22] used the SVM method with spectral roll-off, slope, minimum, median, coefficient of variation, inverse coefficient of variation, trimmed mean, skewness, kurtosis, and $1^{st}$, $57^{th}$, $95^{th}$ and $99^{th}$ percentiles as features, reporting an accuracy higher than 90% for the recognition of some environments, such as restaurant, casino, playground, street traffic, street with ambulance, train, nature at day time, nature at night time, ocean, and river.

In [23], the Linear Discriminant Classifier (LDC) was used with microphone data in order to recognize several ADL, including eating, drinking, clearing the throat, relaxing, laughing, coughing, sniffling, and talking, using several features, including log power, total Root-Mean-Square (RMS) energy, spectral centroid, spectral flux, spectral variance, spectral skewness, spectral kurtosis, spectral slope, spectral roll-off, zero crossing rate, MFCC, minimum, maximum, mean, RMS, median, $1^{st}$ and $3^{rd}$ quartiles, interquartile range, standard deviation, skewness, kurtosis, number of peaks, mean distance of peaks, mean amplitude of peaks, mean crossing rate and linear regression slope, where the best reported accuracy was achieved using the total RMS energy, spectral centroid, spectral flux, spectral variance, spectral skewness, spectral kurtosis, spectral slope, spectral roll-off and MFCC as features and the average of the reported accuracy was 66.5%.

The Artificial Neural Networks (ANN) is one of the most methods used for the recognition of the ADL and their environments using the acoustic signals. In [24], the authors implemented ANN method (*i.e.,* Multi-Layer Perceptron (MLP)) with MFCC as features for the recognition of acoustic warning signals of emergency vehicles (police, fire department and ambulance), reporting an highest accuracy of 96.7%.

Another study [25] uses ANN for the recognition of boll impact, metal impact, wood impact, plastic impact, door opening/closing, typing, knocking, telephone ringing, grains falling, spray and whistle, using time-variance and frequency-variance patterns as features, reporting an average accuracy of 92%.

In [26], the ANN was used for the recognition of sneezing, dog barking, clock ticking, baby crying, crowing rooster, raining, sound of sea waves, fire crackling, sound of helicopter, and sound of chainsaw with some features, such as zero crossing rate, MFCC, spectral flatness, and spectral centroid, reporting an accuracy around 74.5%.

Another type of ANN named as Feedforward neural networks was used in [27] for the recognition of sirens from emergency vehicles, car horns, and normal street sounds with MFCC and zero crossing rate as features, reporting an accuracy between 70% and 90%.

Deep Neural Networks (DNN) is another type of neural networks used for the recognition of laughing, singing, crying, arguing and sighing with MFCC as features, reporting results with reliable accuracy [28]. The authors of [29] also used DNN for the ambient scene analysis (*i.e.,* voice, music, water and traffic), stress detection, emotion recognition, and speaker identification with MFCC as features, reporting an accuracy between 60% and 90%.

The SVM is another method used for the recognition of ADL and their environments using acoustic signals. In [30], the authors implemented the SVM method for the recognition of keystrokes with MFCC as features, reporting an accuracy of 78.4%.

The authors of [31] used the SVM method for the recognition of several sounds, including beach, crowd football, shaver, birds, dishwasher, sink, brushing teeth, dog, speech, bus, forest, street, car, phone ringing, chair, train station, vacuum cleaner, coffee machine, raining, washing machine, computer keyboard and restaurant, using MFCC as features and reporting an accuracy around 80%. The SVM method is also used for the recognition of sleeping using MFCC and sound pressure level (SPL) as features, reporting accuracies between 75% and 81% [32, 33].

The Hidden Markov models (HMM) is another method used for the recognition of ADL and their environments using acoustic signals. In [34], the authors used HMM for the recognition of several sounds, such as car, truck, moped, aircraft, and train, using computation and storage of noise levels, one-third-octave spectra, statistical indices, and detection of noise events based on thresholds as features, reporting an accuracy higher than 95%. In [35], the authors recognized the idle state and the cicada singing sounds with HMM, based on the frequency bands and ratio, reporting results with a reliable accuracy.

The Gaussian Mixture Model (GMM) is another method used for the recognition of ADL and their environments using acoustic signals. In [36], the authors used GMM with MFCC as features for the recognition of calls during driving, reporting an accuracy around 86%. On the other hand, the authors of [37] used GMM with zero crossing rate, RMS, MFCC, and low energy frame rate as features for the recognition of emotional states, reporting an accuracy between 65% and 100%.

The authors of [38] used Random Forests and SVM methods for the recognition of air conditioner, car horn, children playing, dog bark, drilling, idling, gum shot, jackhammer, siren, and street music sounds, using MFCC and motif features, reporting an accuracy between 26.45% and 55.68% with SVM, and between 70.55% and 85% with Random Forests.

In [39], the authors used the decision tree and HMM methods for the recognition of several ADL and environments, including reading, meeting, chatting, assisting conference talks, lectures, music, driving, elevator, walking, airplane, fan, vacuuming, shower, clapping, raining, climbing stairs, and wind, using zero crossing rate, low energy frame rate, spectral flux, spectral roll-off, bandwidth, normalized weighted phase deviation, and Relative Spectral Entropy (RSE), reporting an accuracy higher than 78%.

The authors of [40] implemented the GMM, Feed-Forward DNN, Recurrent Neural Networks (RNN), and SVM for the recognition of baby crying and smoking alarm, using MFCC, spectral centroid, spectral flatness, spectral roll-off, spectral kurtosis, and zero crossing rate, reporting accuracies between 2% and 24%.

The SVM, diverse density (DD), and expected maximization (EM) methods were implemented in [41] for the recognition of several sounds, including cutlery, water, voice, ambient, and music, using MFCC, spectral flux, spectral centroid, bandwidth, Normalized Mel-Frequency Bands, zero crossing rate, and low energy frame rate as features, reporting an average accuracy of 87%.

In [42], several sounds were identified, including coffee machine brewing, hand washing, walking, elevator, door opening/closing, and silence, using k-Nearest Neighbour (k-NN), SVM and GMM methods with some features, such as zero crossing rate, short-time energy, temporal centroid, energy entropy, autocorrelation, RMS, spectral centroid, spectral spread, spectral roll-off point, spectral flux, spectral entropy, and MFCC methods. The highest accuracies achieved with the different methods are 97.9%, with k-NN, 90%, with GMM, and 100%, with SVM [42].

The authors of [43] implemented the Random Forest, HMM, GMM, SVM, ANN, k-NN, and deep belief network methods in order to recognize babble, driving, machinery, crowded restaurant, street, air conditioner, washer, dryer, and vacuum cleaner, with MFCC, band periodicity, and band entropy, reporting results with a reliable accuracy.

In [44], the authors implemented Naïve Bayes, k-NN, Random Forest, and Bayesian Networks methods for the recognition of several nursing activities, including measurement of height, patient sitting, assisting doctor, attaching/measuring/removing electrocardiography (ECG), changing bandage, cleaning body, examining edema, and washing hands, using several features, including mean of intensity, mean, variance of intensity, variance, mean of Fast Fourier Transform (FFT)-domain energy, and covariance between intensities. The results reported are 56.10%, with k-NN and Naïve Bayes, 73.18%, with k-NN and Bayesian Networks, 55.15%, with Naïve Bayes only, 80.96%, with Naïve Bayes and Bayesian Networks, 59.03%, with Random Forest and Naïve Bayes, and 67.83%, with Random Forest and Bayesian Networks [44].

The authors of [45] recognized several sounds, including alarms, birds, clapping, dogs, footsteps, motorcycles, raining, rivers, sea waves, and wind, using k-NN, Naïve Bayes, SVM, C4.5 decision tree, logistic regression, and ANN, imputing several features, including zero crossing rate, skewness, kurtosis, spectral centroid, spectral spread, spectral flux, spectral slope, spectral roll-off, spectral skewness, spectral kurtosis, spectral flatness measure, spectral crest factor, spectral sharpness, Chroma vectors, spectral smoothness, spectral variability, and MFCC. The highest reported accuracies are 45%, with k-NN, 45%, with Naïve Bayes, 54%, with SVM, 45%, with C4.5 decision tree, 44%, with logistic regression, and 54%, with ANN [45].

In [46], a fall detection method was developed with k-NN, SVM, least squares method (LSM), and ANN methods with spectrogram, MFCC, linear predictive coding (LPC), and matching pursuit (MP) as features, reporting 98% of accuracy.

Following the research studies available in the literature, the table 1 shows the ADL and environments recognized with the use of the microphone, verifying that the standing activities are well differentiated with acoustic data.

*Table 1 - Distribution of the ADL and environments extracted in the studies analyzed*

| ADL: | Number of Studies: |
|---|---|
| street with emergency vehicles (police, fire department and ambulance) | 6 |
| sleeping | 5 |
| walking | 5 |
| standing | 5 |
| street traffic | 5 |
| ocean | 5 |
| driving | 4 |
| river | 4 |
| sitting | 3 |
| cleaning with a vacuum cleaner | 3 |
| train | 3 |
| nature | 3 |
| typing | 3 |
| dog barking | 3 |
| baby crying | 3 |
| raining | 3 |
| music | 3 |
| running | 2 |
| lying | 2 |
| going upstairs | 2 |
| going downstairs | 2 |
| drinking | 2 |
| shopping | 2 |
| travelling by car | 2 |
| cooking | 2 |
| watching television | 2 |
| eating | 2 |
| working on a computer | 2 |
| reading | 2 |
| washing dishes | 2 |
| restaurant | 2 |
| laughing | 2 |
| door opening/closing | 2 |
| telephone ringing | 2 |
| helicopter | 2 |
| speech | 2 |
| coffee machine | 2 |
| elevator | 2 |
| social interaction activities | 1 |
| jogging | 1 |
| cycling | 1 |
| cleaning table | 1 |
| going to toilet | 1 |
| waiting in a queue | 1 |
| being a bar | 1 |
| casino | 1 |
| playground | 1 |
| clearing the throat | 1 |

| ADL: | Number of Studies: |
|---|---|
| relaxing | 1 |
| coughing | 1 |
| sniffling | 1 |
| talking | 1 |
| grains falling | 1 |
| whistle | 1 |
| sneezing | 1 |
| clock ticking | 1 |
| arguing | 1 |
| football | 1 |
| shaver | 1 |
| bird | 1 |
| dishwasher | 1 |
| brushing teeth | 1 |
| bus | 1 |
| calling | 1 |
| air conditioner | 1 |
| car horn | 1 |
| children playing | 1 |
| drilling | 1 |
| meeting | 1 |
| chatting | 1 |
| shower | 1 |
| clapping | 1 |
| smoking alarm | 1 |
| hand washing | 1 |

Several features, presented in the table 2, have been used for the recognition of ADL and environments based on acoustic data, showing that the MFCC, zero crossing rate, spectral roll-off, spectral centroid, spectral flux, total RMS energy, mean, standard deviation, minimum, median, and low energy frame rate are the most used features, with more relevance for MFCC.

*Table 2 - Distribution of the features extracted in the studies analyzed*

| Features: | Number of Studies: |
|---|---|
| Mel-Frequency Cepstral Coefficients (MFCC) | 18 |
| zero-crossing rate | 8 |
| spectral roll-off | 6 |
| spectral centroid | 5 |
| spectral flux | 5 |
| total Root-Mean-Square (RMS) energy | 4 |
| mean | 3 |
| standard deviation | 3 |
| minimum | 3 |
| median | 3 |
| low energy frame rate | 3 |
| spectral spread | 2 |
| log power | 2 |
| skewness | 2 |
| kurtosis | 2 |
| sound pressure level (SPL) | 2 |
| bandwidth | 2 |
| Relative Spectral Entropy (RSE) | 2 |
| total spectrum power | 1 |

| Features:                                      | Number of Studies: |
|------------------------------------------------|--------------------|
| sub-band powers                                | 1                  |
| range                                          | 1                  |
| angular degree                                 | 1                  |
| slope                                          | 1                  |
| coefficient of variation                       | 1                  |
| inverse coefficient of variation               | 1                  |
| trimmed mean                                   | 1                  |
| percentiles (1st, 57th, 95th, and 99th)        | 1                  |
| spectral variance                              | 1                  |
| spectral skewness                              | 1                  |
| spectral kurtosis                              | 1                  |
| spectral slope                                 | 1                  |
| maximum                                        | 1                  |
| quartiles (1st and 3rd)                        | 1                  |
| interquartile range                            | 1                  |
| number of peaks                                | 1                  |
| mean distance of peaks                         | 1                  |
| mean amplitude of peaks                        | 1                  |
| mean crossing rate                             | 1                  |
| linear regression slope                        | 1                  |
| spectral flatness                              | 1                  |
| threshold                                      | 1                  |
| noise level                                    | 1                  |
| one-third-octave spectra                       | 1                  |
| statistical indices                            | 1                  |
| motif                                          | 1                  |
| normalized weighted phase deviation            | 1                  |
| Normalized Mel-Frequency Bands                 | 1                  |
| short-time energy                              | 1                  |
| temporal centroid                              | 1                  |
| energy entropy                                 | 1                  |
| autocorrelation                                | 1                  |
| spectral entropy                               | 1                  |

At the end, the recognition of ADL and environments may be performed with several methods presented in the table 3, concluding that the most used methods are SVM, MLP, GMM, and DNN methods. Following the most used methods for the recognition of ADL and environments using the acoustic signal, implemented in more than 3 studies analyzed, the method that reports the best average accuracy in the recognition of ADL and environments is the MLP, with an average accuracy of 88%.

*Table 3 - Distribution of the classification methods used in the studies analyzed*

| Methods:                                                     | Number of Studies: | Average of Reported Accuracy: |
|--------------------------------------------------------------|--------------------|-------------------------------|
| Support Vector Machine (SVM)                                 | 10                 | 77%                           |
| Gaussian mixture model (GMM)                                 | 5                  | 76%                           |
| Artificial Neural Networks (ANN) / Multi-Layer Perceptron (MLP) | 3               | 88%                           |
| Deep Neural Networks (DNN)                                   | 3                  | 68%                           |
| Hidden Markov Models (HMM)                                   | 2                  | 87%                           |
| J48 decision tree                                            | 2                  | 84%                           |
| FT decision tree                                             | 2                  | 84%                           |
| LMT decision tree                                            | 2                  | 84%                           |
| k-Nearest Neighbour (k-NN)                                   | 1                  | 98%                           |
| Gradient Boosting Decision Tree                              | 1                  | 92%                           |

| Methods: | Number of Studies: | Average of Reported Accuracy: |
|---|---|---|
| IBk lazy algorithm | 1 | 91% |
| logistic regression | 1 | 90% |
| linear regression | 1 | 90% |
| Feedforward neural networks | 1 | 90% |
| diverse density (DD) | 1 | 87% |
| expected maximization (EM) | 1 | 87% |
| Random Forests | 1 | 85% |
| Linear Discriminant Classifier (LDC) | 1 | 67% |
| Recurrent Neural Networks (RNN) | 1 | 24% |

## 3. Methods

In coherence with the methods defined in the previous studies [4, 14] for the development of the framework for the recognition of ADL and their environments [5-7], the methods developed in this study should be separated in several methods, such as data acquisition, data processing, data fusion, and artificial intelligence methods, where the fusion of the data and the application of artificial intelligence methods are performed at the same time.

Following the steps for the creation of the method, firstly, the section 3.1 presented the data acquisition methods. Secondly, the data processing methods are presented in the section 3.2. Finalizing this section with the presentation of the data fusion and artificial intelligence methods in the section 3.3.

### 3.1. Data Acquisition

The data acquisition was performed with a mobile application installed in a BQ Aquarius device [47] with the Android operating system [48, 49] installed, which allows the captures of the sensors' data and, at the first stage, saves the data acquired from the microphone in a raw format into text files, and, in a second stage, the data captured from the accelerometer, magnetometer, and gyroscope sensors are also saved in text files. The sensors' data is captured in slots of 5 seconds every 5 minutes in background and a frequency of data acquisition by the accelerometer, magnetometer, and gyroscope sensors is around 10ms. Before the experiments, the user selected the ADL that are performed and/or the environments were the ADL are performed. The experiments were performed with the mobile device in the pocket by people aged between 16 and 60 years old and different lifestyles. Following the most common environments and the most identified ADL in the literature, the allowed ADL in the mobile application are running, walking, going upstairs, sleeping, going downstairs, and standing, and the allowed environments in the mobile application are bar, classroom, gym, kitchen, library, street, hall, watching TV and bedroom. A minimum of 2000 experiments for each ADL and environment have been acquired and stored in the ALLab MediaWiki [50].

### 3.2. Data Processing

Another module of the framework for the recognition of ADL and their environments is composed by data processing methods. This module is composed by data cleaning methods, presented in the section 3.2.1, and methods for extraction of feature, presented in the section 3.2.2.

#### 3.2.1. Data Cleaning

Data cleaning methods are different for each type of sensors, removing the noise and the invalid data present in the data acquired. For the data captured with the microphone available in the mobile device, the Fast Fourier Transform (FFT) [51] is the best method to apply for the extraction of the frequencies of the audio signal, handling the reduction of the environmental noise. For the data captured with the accelerometer, magnetometer, and gyroscope sensors, the low pass filter [52] is the best method for the reduction of the noise captured during the ADL with movement.

#### 3.2.2. Feature Extraction

In coherence with our previous studies [4, 14], based on the data filtered and the most features extracted in the studies available in the literature, the features extracted for the methods using acoustic data for the recognition of environments were the 26 MFCC coefficients, the Standard Deviation of the raw signal, the Average of the raw signal, the Maximum value of the raw signal, the Minimum value of the raw signal, the Variance of the of the raw signal, and the Median of the raw signal. On the other hand, the features extracted from the accelerometer, gyroscope, and magnetometer sensors were the 5 greatest distances between the maximum peaks, the Average of the maximum peaks, the Standard Deviation of the maximum peaks, the Variance of the maximum peaks, the Median of the maximum peaks, the Standard Deviation of the raw signal, the Average of the raw signal, the Maximum value of the raw signal, the Minimum value of the raw signal, the Variance of the of the raw signal, the Median of the raw signal, and the environment recognized.

### 3.3. Identification of Activities of Daily Living and their environment

In continuation of our previous studies [4, 14] using the accelerometer, gyroscope and magnetometer sensors, this study creates datasets with the features extracted from the acoustic data for the recognition of the environment (section 3.3.1), the features extracted from the fusion of the accelerometer data and the environment recognized (section 3.3.2), the features extracted from the fusion of the accelerometer and magnetometer data and the environment recognized (section 3.3.3), and the features extracted from the fusion of the accelerometer, magnetometer and gyroscope data and the environment recognized (section 3.3.4). At the end of this section, the artificial intelligence methods for the recognition of ADL and their environments are presented in the section 3.3.5.

#### 3.3.1. Identification of environments of Activities of Daily Living using Microphone

Regarding the features extracted from each environment, four datasets have been constructed with features extracted from the microphone data acquired in the defined environments, having 2000 records from each environment. The datasets defined are:
- **Dataset 1:** Composed by 26 MFCC coefficients, Standard Deviation of the raw signal, Average of the raw signal, Maximum value of the raw signal, Minimum value of the raw signal, Variance of the of the raw signal, and Median of the raw signal, extracted from the microphone data;
- **Dataset 2:** Composed by Standard Deviation of the raw signal, Average of the raw signal, Maximum value of the raw signal, Minimum value of the raw signal, Variance of the of the raw signal, and Median of the raw signal, extracted from the microphone data;
- **Dataset 3:** Composed by Standard Deviation of the raw signal, Average of the raw signal, Variance of the of the raw signal, and Median of the raw signal, extracted from the microphone data;
- **Dataset 4:** Composed by Standard Deviation of the raw signal, and Average of the raw signal, extracted from the microphone data.

#### 3.3.2. Data fusion of the environment recognized with the Accelerometer data for the recognition of standing activities

Regarding the features extracted from each standing activity, five datasets have been constructed with features extracted from the accelerometer data acquired during the performance of the two standing activities, having 2000 records from each activity. This method allows the distinction between sleeping and watching TV. The datasets defined are:
- **Dataset 1:** Composed by 5 greatest distances between the maximum peaks, Average of the maximum peaks, Standard Deviation of the maximum peaks, Variance of the maximum peaks, Median of the maximum peaks, Standard Deviation of the raw signal, Average of the raw signal, Maximum value of the raw signal, Minimum value of the raw signal, Variance of the of the raw signal, and Median of the raw signal, extracted from the accelerometer sensor, and the environment recognized with the features defined in the section 3.3.1;

- **Dataset 2:** Composed by Average of the maximum peaks, Standard Deviation of the maximum peaks, Variance of the maximum peaks, Median of the maximum peaks, Standard Deviation of the raw signal, Average of the raw signal, Maximum value of the raw signal, Minimum value of the raw signal, Variance of the of the raw signal, and Median of the raw signal, extracted from the accelerometer sensor, and the environment recognized with the features defined in the section 3.3.1;
- **Dataset 3:** Composed by Standard Deviation of the raw signal, Average of the raw signal, Maximum value of the raw signal, Minimum value of the raw signal, Variance of the of the raw signal, and Median of the raw signal, extracted from the accelerometer sensor, and the environment recognized with the features defined in the section 3.3.1;
- **Dataset 4:** Composed by Standard Deviation of the raw signal, Average of the raw signal, Variance of the of the raw signal, and Median of the raw signal, extracted from the accelerometer sensor, and the environment recognized with the features defined in the section 3.3.1;
- **Dataset 5:** Composed by Standard Deviation of the raw signal, and Average of the raw signal, extracted from the accelerometer sensor, and the environment recognized with the features defined in the section 3.3.1.

### 3.3.3. Data fusion of the environment recognized with the Accelerometer and Magnetometer data for the recognition of standing activities

Regarding the features extracted from each standing activity, five datasets have been constructed with features extracted from the accelerometer and magnetometer sensors' data acquired during the performance of the two standing activities, having 2000 records from each activity. This method allows the distinction between sleeping and watching TV. The datasets defined are:

- **Dataset 1:** Composed by 5 greatest distances between the maximum peaks, Average of the maximum peaks, Standard Deviation of the maximum peaks, Variance of the maximum peaks, Median of the maximum peaks, Standard Deviation of the raw signal, Average of the raw signal, Maximum value of the raw signal, Minimum value of the raw signal, Variance of the of the raw signal, and Median of the raw signal, extracted from the accelerometer and magnetometer sensors, and the environment recognized with the features defined in the section 3.3.1;
- **Dataset 2:** Composed by Average of the maximum peaks, Standard Deviation of the maximum peaks, Variance of the maximum peaks, Median of the maximum peaks, Standard Deviation of the raw signal, Average of the raw signal, Maximum value of the raw signal, Minimum value of the raw signal, Variance of the of the raw signal, and Median of the raw signal, extracted from the accelerometer and magnetometer sensors, and the environment recognized with the features defined in the section 3.3.1;
- **Dataset 3:** Composed by Standard Deviation of the raw signal, Average of the raw signal, Maximum value of the raw signal, Minimum value of the raw signal, Variance of the of the raw signal, and Median of the raw signal, extracted from the accelerometer and magnetometer sensors, and the environment recognized with the features defined in the section 3.3.1;
- **Dataset 4:** Composed by Standard Deviation of the raw signal, Average of the raw signal, Variance of the of the raw signal, and Median of the raw signal, extracted from the accelerometer and magnetometer sensors, and the environment recognized with the features defined in the section 3.3.1;
- **Dataset 5:** Composed by Standard Deviation of the raw signal, and Average of the raw signal, extracted from the accelerometer and magnetometer sensors, and the environment recognized with the features defined in the section 3.3.1.

### 3.3.4. Data fusion of the environment recognized with the Accelerometer, Magnetometer and Gyroscope data for the recognition of standing activities

Regarding the features extracted from each standing activity, five datasets have been constructed with features extracted from the accelerometer, magnetometer and gyroscope sensors' data acquired during the performance of the two standing activities, having 2000 records from each activity. This method allows the distinction between sleeping and watching TV. The datasets defined are:
- **Dataset 1:** Composed by 5 greatest distances between the maximum peaks, Average of the maximum peaks, Standard Deviation of the maximum peaks, Variance of the maximum peaks, Median of the maximum peaks, Standard Deviation of the raw signal, Average of the raw signal, Maximum value of the raw signal, Minimum value of the raw signal, Variance of the of the raw signal, and Median of the raw signal, extracted from the accelerometer, magnetometer and gyroscope sensors, and the environment recognized with the features defined in the section 3.3.1;
- **Dataset 2:** Composed by Average of the maximum peaks, Standard Deviation of the maximum peaks, Variance of the maximum peaks, Median of the maximum peaks, Standard Deviation of the raw signal, Average of the raw signal, Maximum value of the raw signal, Minimum value of the raw signal, Variance of the of the raw signal, and Median of the raw signal, extracted from the accelerometer, magnetometer and gyroscope sensors, and the environment recognized with the features defined in the section 3.3.1;
- **Dataset 3:** Composed by Standard Deviation of the raw signal, Average of the raw signal, Maximum value of the raw signal, Minimum value of the raw signal, Variance of the of the raw signal, and Median of the raw signal, extracted from the accelerometer, magnetometer and gyroscope sensors, and the environment recognized with the features defined in the section 3.3.1;
- **Dataset 4:** Composed by Standard Deviation of the raw signal, Average of the raw signal, Variance of the of the raw signal, and Median of the raw signal, extracted from the accelerometer, magnetometer and gyroscope sensors, and the environment recognized with the features defined in the section 3.3.1;
- **Dataset 5:** Composed by Standard Deviation of the raw signal, and Average of the raw signal, extracted from the accelerometer, magnetometer and gyroscope sensors, and the environment recognized with the features defined in the section 3.3.1.

### 3.3.5. Artificial Intelligence

Based on the literature review related to the use of acoustic data for the recognition of the environments, presented in the section 2, one of the most used methods for the recognition of environments is the ANN, reporting better accuracy than other most used methods, such as SVM, GMM, and DNN. In addition, based on the literature reviews about the recognition of ADL using accelerometer, magnetometer and gyroscope sensors, presented in our previous studies [4, 14], one of the most used methods for the recognition of ADL is the ANN, reporting better accuracy than SVM, KNN, Random Forest, and Naïve Bayes, however the results obtained in these studies [4, 14] proved that DNN reports better accuracy.

For the identification of the best methods for the recognition of environments and standing activities proposed in the sections 3.3.1, 3.3.2, 3.3.3 and 3.3.4, this study explores the use of three types of neural networks, such as MLP, Feedforward Neural Network, and DNN, with different frameworks, these are:
- MLP with Backpropagation, applied with Neuroph framework [15];
- Feedforward Neural Network with Backpropagation, applied with Encog framework [16];
- Deep Neural Networks, applied with DeepLearning4j framework [17].

Before the implementation of the MLP with Backpropagation, and the Feedforward Neural Network with Backpropagation, the datasets should be normalized with the MIN/MAX normalizer [53], implementing these methods with non-normalized and normalized data to verify if the normalization increases the accuracy of the recognition of ADL and environments.

On the other hand, before the implementation of DNN method, the datasets should be normalized with mean and standard deviation [54] and applied the $L_2$ regularization [55], implementing this method with non-normalized and normalized data to verify if the normalization increases the accuracy of the recognition of ADL and environments.

The number of training iterations is another factor that may affect the accuracy of the methods and we defined three limits for the verification of the best number of iterations for the recognition of ADL and environments, these are 1M, 2M and 4M.

Based on the datasets defined in the sections 3.3.1, 3.3.2, 3.3.3 and 3.3.4, the created methods should be implemented in a framework for the recognition of ADL and environments defined in [5-7]. For the recognition of common ADL, as concluded in the previous studies [4, 14], the method that should be implemented is DNN with normalized data and $L_2$ regularization, however this research will identify the best method for the recognition of the environments, based in the datasets defined in the section 3.3.1, and the best methods fort the distinction between standing activities, based on the datasets defined in the section 3.3.2, 3.3.3 and 3.3.4.

## 4. Results

The results of this paper are focused on the creation of one method for the recognition of the environments using the microphone data, and three methods for the recognition of standing activities with different number of sensors. Firstly, the results of the creation of a method for the recognition of the environment are presented in the section 4.1. Secondly, the results of the creation of a method with accelerometer sensor are presented in the section 4.2. Thirdly, the results of the creation of a method with accelerometer and magnetometer sensors are presented in the section 4.3. Finally, the results of the creation of a method with accelerometer, magnetometer, and gyroscope sensors are presented in the section 4.4.

### 4.1. Identification of the environment of the Activities of Daily Living with Microphone

Based on the datasets defined in the section 3.3.1, the three types of neural networks proposed in the section 3.3.5 were implemented, these are MLP with Backpropagation, Feedforward Neural Network with Backpropagation, and DNN. The datasets defined for training and testing phases are composed by 16000 records, where each environments has 2000 records.

Firstly, the results of the implementation of the MLP with Backpropagation using the Neuroph framework are presented in the figure 1, verifying that the results have very low accuracy with all datasets. With non-normalized data (figure 1-a), the results achieved are between 10% and 15%. And, with normalized data (figure 1-b), the results obtained are between 10% and 20%, where the best results are achieved with dataset 1.

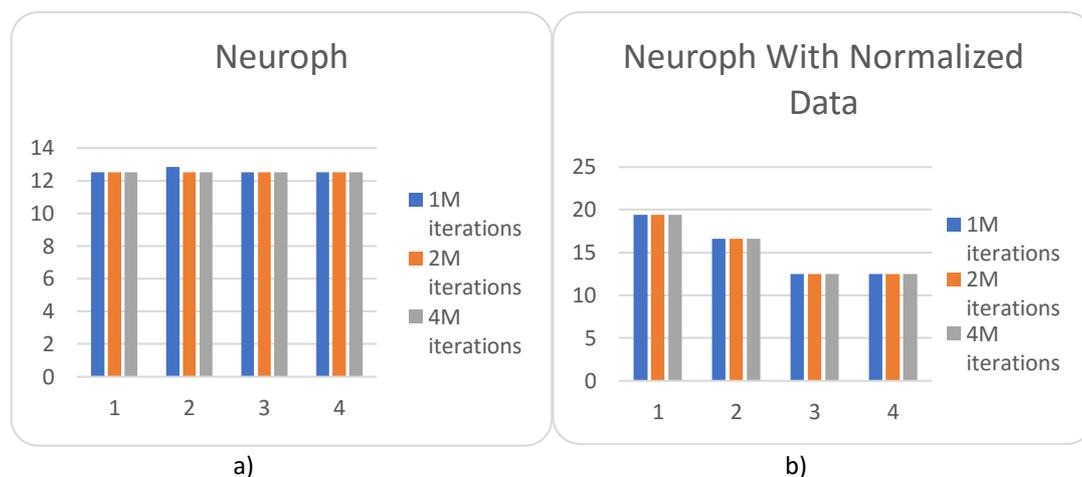

*Figure 1 –Results obtained with Neuroph framework for the different datasets of microphone data (horizontal axis) and different maximum number of iterations (series), obtaining the accuracy in percentage (vertical axis). The figure a) shows the results with data without normalization. The figure b) shows the results with normalized data.*

Secondly, the results of the implementation of the Feedforward Neural Network with Backpropagation using the Encog framework are presented in the figure 2. In general, this type of neural network reports better results with non-normalized data. With non-normalized data (figure 2-a), the neural networks reports results higher than 70% with dataset 1 with all maximum number of training iterations, dataset 2 with 1M of training iterations, and dataset 4 with 4M of training iterations. With normalized data (figure2-b), the neural networks reports results below than 60%, but the results achieved are higher than 60% with the dataset 4 trained over 1M and 2M of iterations.

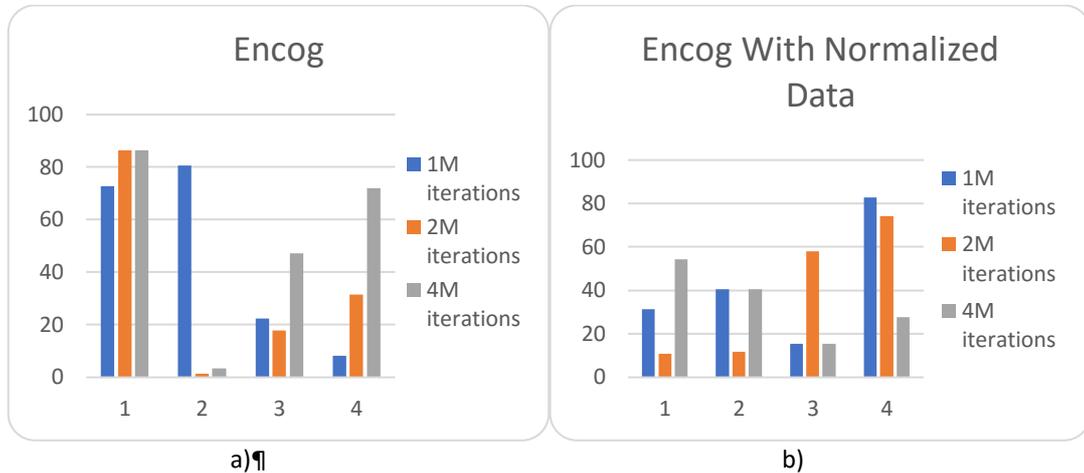

*Figure 2 –Results obtained with Encog framework for the different datasets of microphone data (horizontal axis) and different maximum number of iterations (series), obtaining the accuracy in percentage (vertical axis). The figure a) shows the results with data without normalization. The figure b) shows the results with normalized data.*

Finally, the results of the implementation of DNN with DeepLearning4j framework are presented in the figure 3. With non-normalized data (figure 3-a), the results obtained are below 20% with datasets 1 and 2, and the results obtained are higher than 40% with datasets 3 and 4. On the other hand, with normalized data (figure 3-b), the results reported are round 50% with all datasets.

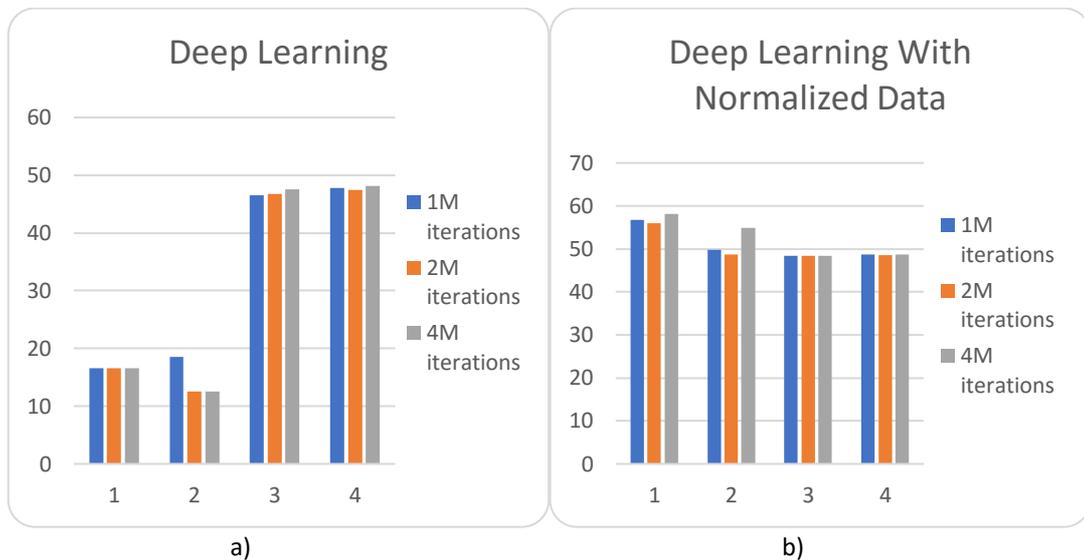

*Figure 3 –Results obtained with DeepLearning4j framework for the different datasets of microphone data (horizontal axis) and different maximum number of iterations (series), obtaining the accuracy in percentage (vertical axis). The figure a) shows the results with data without normalization. The figure b) shows the results with normalized data.*

In table 4, the maximum accuracies achieved with the different types of neural networks are related with the different datasets used for the microphone data, and the maximum number of training iterations, verifying that the best results are achieved with the Feedforward Neural Network with Backpropagation with non-normalized data.

*Table 4 - Best accuracies obtained with the different frameworks, datasets and number of iterations for the recognition of environments using microphone data.*

| | FRAMEWORK | DATASETS | ITERATIONS NEEDED FOR TRAINING | BEST ACCURACY ACHIEVED (%) |
|---|---|---|---|---|
| **NOT NORMALIZED DATA** | NEUROPH | 2 | 1M | 12.86 |
| | ENCOG | 1 | 2M | 86.50 |
| | DEEP LEARNING | 4 | 4M | 48.11 |
| **NORMALIZED DATA** | NEUROPH | 1 | 1M | 19.43 |
| | ENCOG | 4 | 1M | 82.75 |
| | DEEP LEARNING | 4 | 4M | 48.74 |

In conclusion, the method for the recognition of the environment that should be implemented in the framework for the recognition of ADL and their environments is the Feedforward Neural Network with Backpropagation using non-normalized data, because achieves results around 86.50% with the dataset 1.

## 4.2. Identification of the standing activities with the environment recognized and the Accelerometer sensor

Based on the datasets defined in the section 3.3.2, the three types of neural networks proposed in the section 3.3.5 were implemented, these are MLP with Backpropagation, Feedforward Neural Network with Backpropagation, and DNN. The datasets defined for training and testing phases are composed by 4000 records, where each ADL has 2000 records.

Firstly, the results of the implementation of the MLP with Backpropagation using the Neuroph framework are presented in the figure 4, verifying that the results have reliable accuracy with all datasets. With non-normalized data (figure 4-a), the results achieved are between 50% and 100%, where the better accuracy was achieved with the datasets 1 and 4. And, with normalized data (figure 4-b), the results obtained are always around 100% with all datasets.

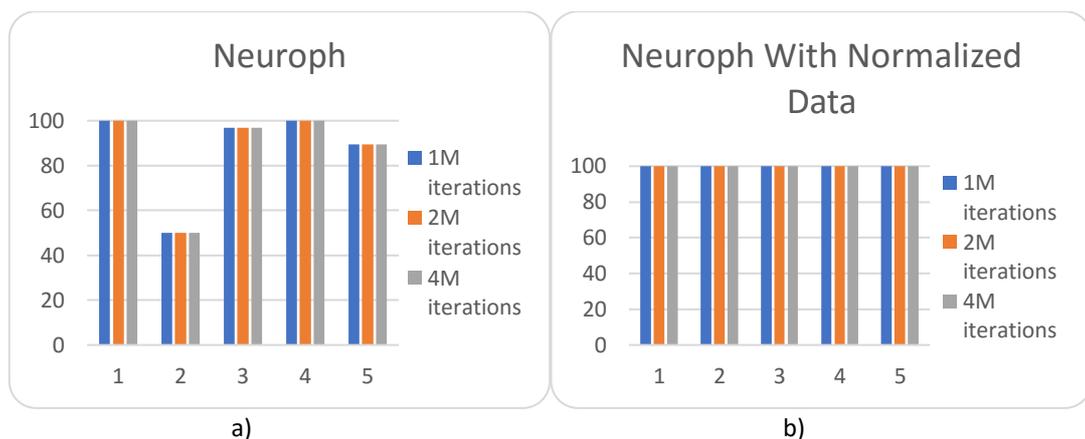

*Figure 4 –Results obtained with Neuroph framework for the different datasets of environment and accelerometer data (horizontal axis) and different maximum number of iterations (series), obtaining the accuracy in percentage (vertical axis). The figure a) shows the results with data without normalization. The figure b) shows the results with normalized data.*

Secondly, the results of the implementation of the Feedforward Neural Network with Backpropagation using the Encog framework are presented in the figure 5, verifying that the results have reliable accuracy with all datasets. With non-normalized data (figure 5-a), the results achieved are around 100%, except with dataset 1 that achieves an accuracy around 50%. And, with normalized data (figure 5-b), the results obtained are always around 100% with all datasets.

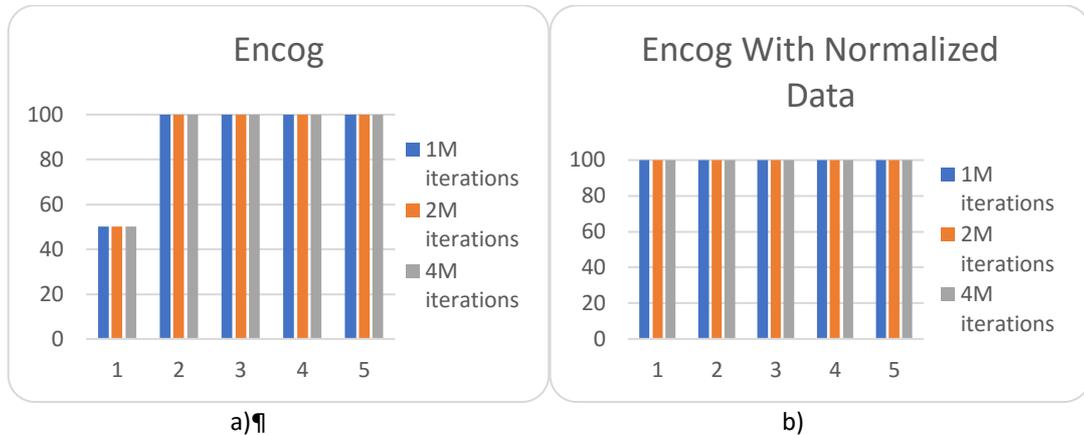

*Figure 5 –Results obtained with Encog framework for the different datasets of environment and accelerometer data (horizontal axis) and different maximum number of iterations (series), obtaining the accuracy in percentage (vertical axis). The figure a) shows the results with data without normalization. The figure b) shows the results with normalized data.*

Finally, the results of the implementation of DNN with DeepLearning4j framework are presented in the figure 6, verifying that the results have reliable accuracy with all datasets. With non-normalized data (figure 6-a), the results obtained are around 100% with datasets 2, 4 and 5 with all training iterations, and with dataset 3 with 4M iterations, but the results obtained with other datasets are below the expectations. On the other hand, with normalized data (figure 6-b), the results obtained are always around 100% with all datasets.

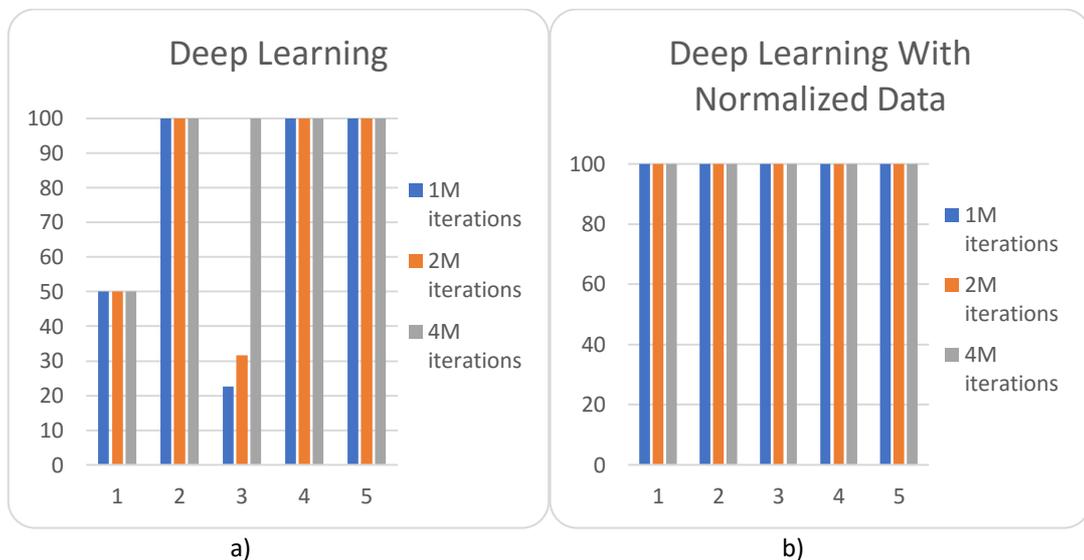

*Figure 6 –Results obtained with DeepLearning4j framework for the different datasets of environment and accelerometer data (horizontal axis) and different maximum number of iterations (series), obtaining the accuracy in percentage (vertical axis). The figure a) shows the results with data without normalization. The figure b) shows the results with normalized data.*

In table 5, the maximum accuracies achieved with the different types of neural networks are presented with the relation of the different datasets used for the environment recognized and the accelerometer data, and the maximum number of iterations, verifying that the use of all neural networks achieves reliable results.

*Table 5 - Best accuracies obtained with the different frameworks, datasets and number of iterations for the recognition of standing activities with the accelerometer data and the environments recognized.*

|  | FRAMEWORK | DATASETS | ITERATIONS NEEDED FOR TRAINING | BEST ACCURACY ACHIEVED (%) |
|---|---|---|---|---|
| **NOT NORMALIZED DATA** | NEUROPH | 1 | 1M | 100.00 |
|  | ENCOG | 2 | 1M | 100.00 |
|  | DEEP LEARNING | 2 | 1M | 100.00 |
| **NORMALIZED DATA** | NEUROPH | 1 | 1M | 100.00 |
|  | ENCOG | 1 | 1M | 100.00 |
|  | DEEP LEARNING | 1 | 1M | 100.00 |

Regarding the results obtained, in the case of the use of the environment recognized and the accelerometer data in the module for the recognition of standing activities in the framework for the identification ADL and their environments, the type of neural networks that should be used is a DNN with normalized data, because the results obtained are always 100%.

## 4.3. Identification of the standing activities with the environment recognized and the Accelerometer and Magnetometer sensors

Based on the datasets defined in the section 3.3.3, the three types of neural networks proposed in the section 3.3.5 were implemented, these are MLP with Backpropagation, Feedforward Neural Network with Backpropagation, and DNN. The datasets defined for training and testing phases are composed by 4000 records, where each ADL has 2000 records.

Firstly, the results of the implementation of the MLP with Backpropagation using the Neuroph framework are presented in the figure 7, verifying that the results have reliable accuracy with all datasets. With non-normalized data (figure 7-a), the results achieved are around 100%, except with the datasets 1 and 5 that achieves an accuracy around 50%. And, with normalized data (figure 7-b), the results obtained are always around 100% with all datasets.

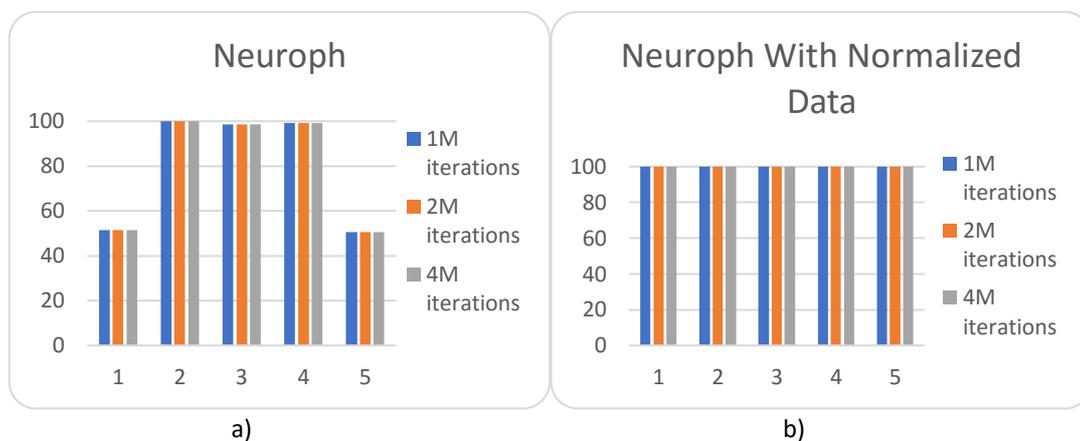

*Figure 7 –Results obtained with Neuroph framework for the different datasets of environment, and accelerometer and magnetometer sensors' data (horizontal axis) and different maximum number of iterations (series), obtaining the accuracy in percentage (vertical axis). The figure a) shows the results with data without normalization. The figure b) shows the results with normalized data.*

Secondly, the results of the implementation of the Feedforward Neural Network with Backpropagation using the Encog framework are presented in the figure 8, verifying that the results have reliable accuracy with all datasets. With non-normalized data (figure 8-a), the results achieved are always around 100%. And, with normalized data (figure 8-b), the results obtained are always around 100% with all datasets.

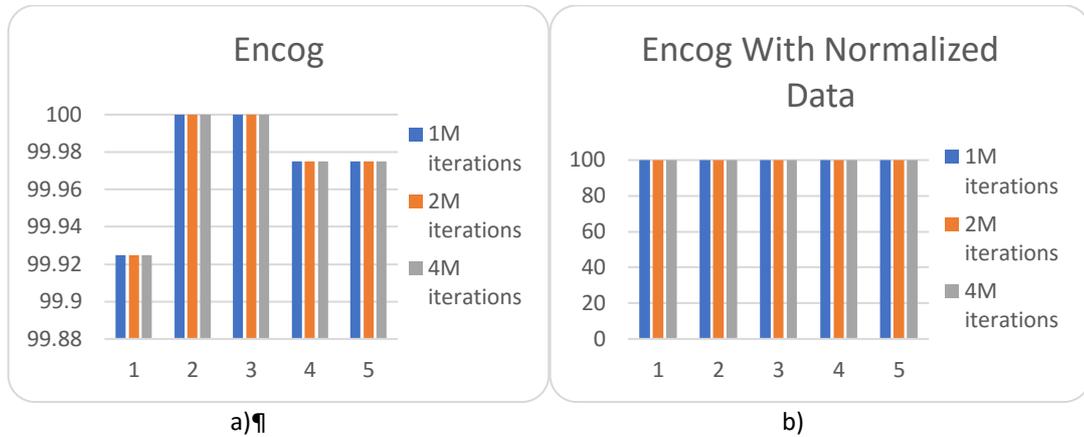

*Figure 8 –Results obtained with Encog framework for the different datasets of environment, and accelerometer and magnetometer sensors' data (horizontal axis) and different maximum number of iterations (series), obtaining the accuracy in percentage (vertical axis). The figure a) shows the results with data without normalization. The figure b) shows the results with normalized data.*

Finally, the results of the implementation of DNN with DeepLearning4j framework are presented in the figure 9. With non-normalized data (figure 9-a), the results obtained are around 100% with dataset 5 with all training iterations, and with dataset 4 with 1M of training iterations, but the results obtained with other datasets are below the expectations. On the other hand, with normalized data (figure 9-b), the results obtained are always around 100% with all datasets.

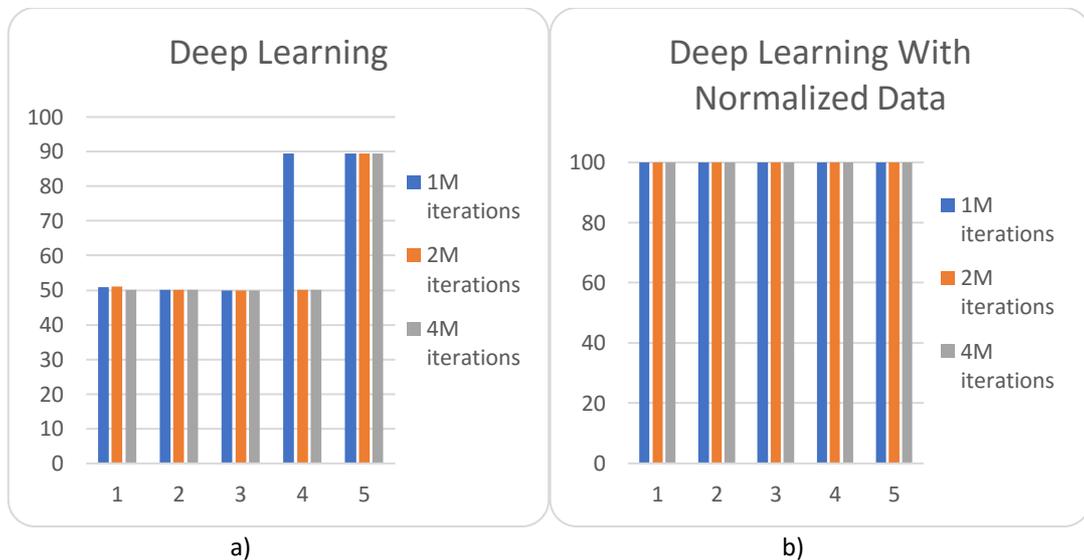

*Figure 9 –Results obtained with DeepLearning4j framework for the different datasets of environment, and accelerometer and magnetometer sensors' data (horizontal axis) and different maximum number of iterations (series), obtaining the accuracy in percentage (vertical axis). The figure a) shows the results with data without normalization. The figure b) shows the results with normalized data.*

In table 6, the maximum accuracies achieved with the different types of neural networks are presented with the relation of the different datasets used for the environment recognized, and the accelerometer and magnetometer sensors' data, and the maximum number of iterations, verifying that the use of all neural networks achieves reliable results.

*Table 6 - Best accuracies obtained with the different frameworks, datasets and number of iterations for the recognition of standing activities with the accelerometer and magnetometer data, and the environments recognized.*

|  | FRAMEWORK | DATASETS | ITERATIONS NEEDED FOR TRAINING | BEST ACCURACY ACHIEVED (%) |
|---|---|---|---|---|
| NOT NORMALIZED DATA | NEUROPH | 4 | 1M | 99.05 |
| | ENCOG | 2 | 1M | 100.00 |
| | DEEP LEARNING | 3 | 1M | 89.55 |
| NORMALIZED DATA | NEUROPH | 1 | 1M | 100.00 |
| | ENCOG | 1 | 1M | 100.00 |
| | DEEP LEARNING | 1 | 1M | 100.00 |

Regarding the results obtained, in the case of the use of the environment recognized, and the accelerometer and magnetometer sensors' data in the module for the recognition of standing activities in the framework for the identification ADL and their environments, the type of neural networks that should be used is a DNN with normalized data, because the results obtained are always 100%.

### 4.4. Identification of the standing activities with the environment recognized and the Accelerometer, Magnetometer and Gyroscope sensors

Based on the datasets defined in the section 3.3.4, the three types of neural networks proposed in the section 3.3.5 were implemented, these are MLP with Backpropagation, Feedforward Neural Network with Backpropagation, and DNN. The datasets defined for training and testing phases are composed by 4000 records, where each ADL has 2000 records.

Firstly, the results of the implementation of the MLP with Backpropagation using the Neuroph framework are presented in the figure 10, verifying that the results have reliable accuracy with all datasets. With non-normalized data (figure 10-a), the results achieved are around 100%, except with the datasets 1 that achieves an accuracy around 50%. And, with normalized data (figure 10-b), the results obtained are always around 100% with all datasets.

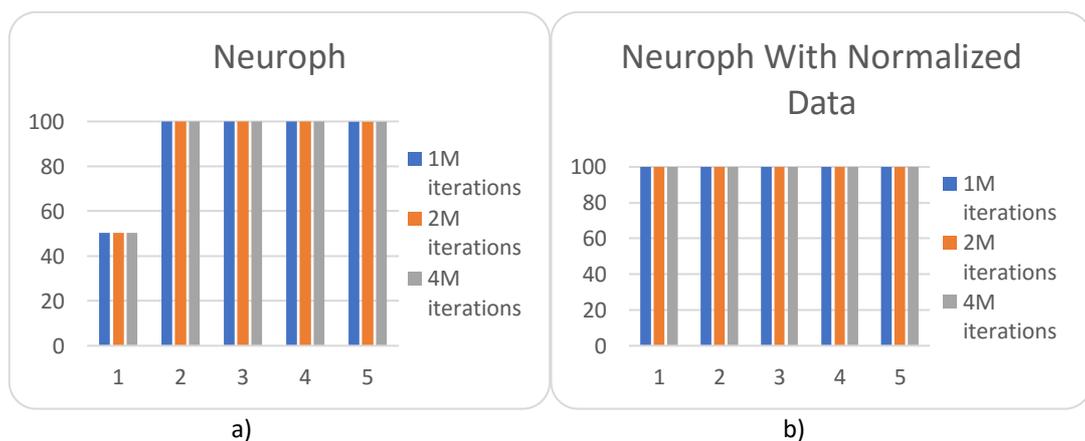

*Figure 10 –Results obtained with Neuroph framework for the different datasets of environment, and accelerometer, magnetometer and gyroscope sensors' data (horizontal axis) and different maximum number of iterations (series), obtaining the accuracy in percentage (vertical axis). The figure a) shows the results with data without normalization. The figure b) shows the results with normalized data.*

Secondly, the results of the implementation of the Feedforward Neural Network with Backpropagation using the Encog framework are presented in the figure 11, verifying that the results have reliable accuracy with all datasets. With non-normalized data (figure 11-a), the results achieved are always around 100%. And, with normalized data (figure 11-b), the results obtained are always around 100% with all datasets.

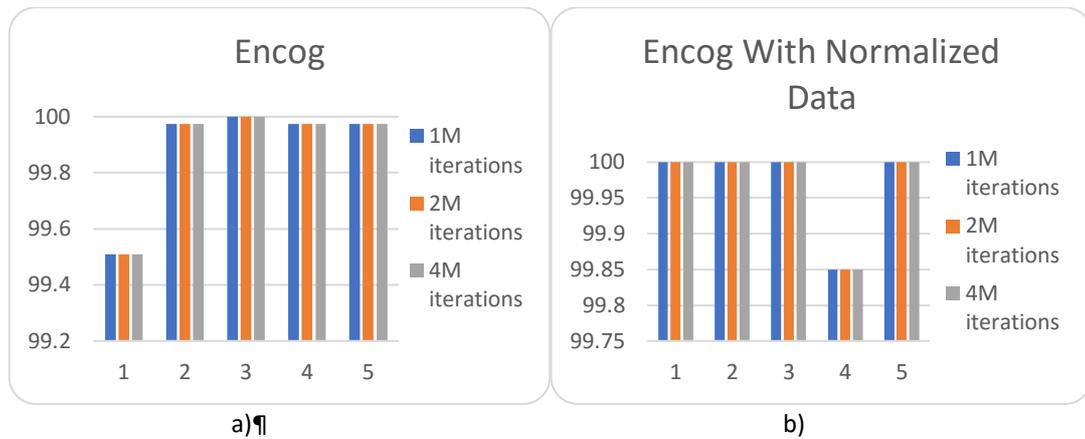

Figure 11 –Results obtained with Encog framework for the different datasets of environment, and accelerometer, magnetometer and gyroscope sensors' data (horizontal axis) and different maximum number of iterations (series), obtaining the accuracy in percentage (vertical axis). The figure a) shows the results with data without normalization. The figure b) shows the results with normalized data.

Finally, the results of the implementation of DNN with DeepLearning4j framework are presented in the figure 12. With non-normalized data (figure 12-a), the results obtained are around 90% with dataset 5 with all training iterations, but the results obtained with other datasets are below the expectations. On the other hand, with normalized data (figure 12-b), the results obtained are always around 100% with all datasets.

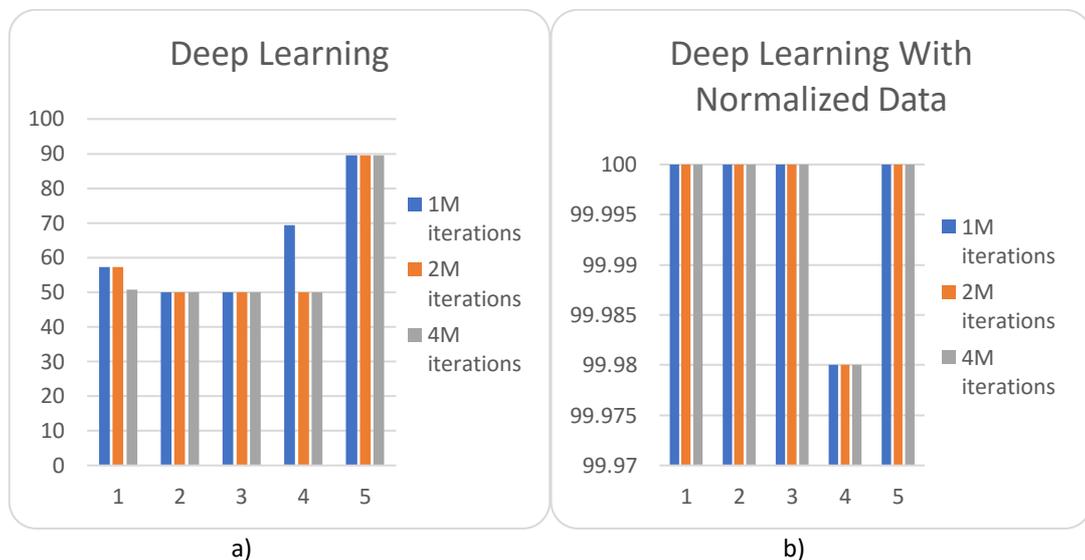

Figure 12 –Results obtained with DeepLearning4j framework for the different datasets of environment, and accelerometer, magnetometer and gyroscope sensors' data (horizontal axis) and different maximum number of iterations (series), obtaining the accuracy in percentage (vertical axis). The figure a) shows the results with data without normalization. The figure b) shows the results with normalized data.

In table 7, the maximum accuracies achieved with the different types of neural networks are presented with the relation of the different datasets used for the environment recognized, and the accelerometer, magnetometer and gyroscope sensors' data, and the maximum number of iterations, verifying that the use of all neural networks achieves reliable results.

Table 7 - Best accuracies obtained with the different frameworks, datasets and number of iterations for the recognition of standing activities with the accelerometer, gyroscope and magnetometer data, and the environments recognized.

| | FRAMEWORK | DATASETS | ITERATIONS NEEDED FOR TRAINING | BEST ACCURACY ACHIEVED (%) |
|---|---|---|---|---|
| NOT NORMALIZED DATA | NEUROPH | 2 | 1M | 100.00 |
| | ENCOG | 3 | 1M | 100.00 |
| | DEEP LEARNING | 5 | 1M | 89.55 |
| NORMALIZED DATA | NEUROPH | 1 | 1M | 100.00 |
| | ENCOG | 1 | 1M | 100.00 |
| | DEEP LEARNING | 1 | 1M | 100.00 |

Regarding the results obtained, in the case of the use of the environment recognized and the accelerometer, magnetometer and gyroscope sensors' data in the module for the recognition of standing activities in the framework for the identification ADL and their environments, the type of neural networks that should be used is a DNN with normalized data, because the results obtained are always 100%.

## 5. Discussion

This research is included in the development of the framework for the recognition of ADL and their environments, presented in [5-7], composed by several modules, including data acquisition, data processing, data fusion, and artificial intelligence methods. The definition of the method for the identification started in the previous studies [4, 14], where several ADL were recognized using accelerometer, gyroscope and magnetometer sensors, these are going downstairs, going upstairs, running, walking and standing with the DNN method, with the normalization of the data and the application of $L_2$ regularization. At the section 3.3.1, the results of the recognition of the environments using the microphone data, where the environments recognized are bar, classroom, gym, kitchen, library, street, hall, watching TV and bedroom with Feedforward neural networks with non-normalized data. Fusing the environment recognized with the accelerometer, gyroscope and magnetometer sensors' data, the recognition of more standing activities (*i.e.,* watching TV and sleeping) was allowed, increasing the number of ADL recognized at this stage of the development of the framework for the recognition of ADL and environments, as presented in the figure 13.

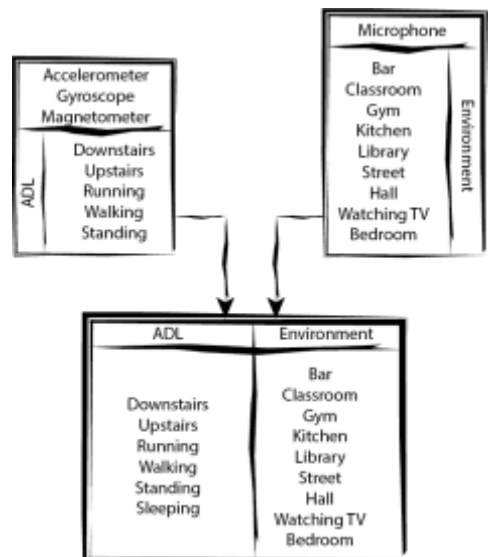

Figure 13 – ADL and environments recognized by the framework for the recognition of ADL and environments.

The choice of the methods for data fusion, and artificial intelligence modules, depends on the number of sensors available on the mobile device, using the maximum number of sensors available on the mobile device, in order to increase the reliability of the method. In the figure 14, a simplified schema for the development of a framework for the identification of ADL is presented.

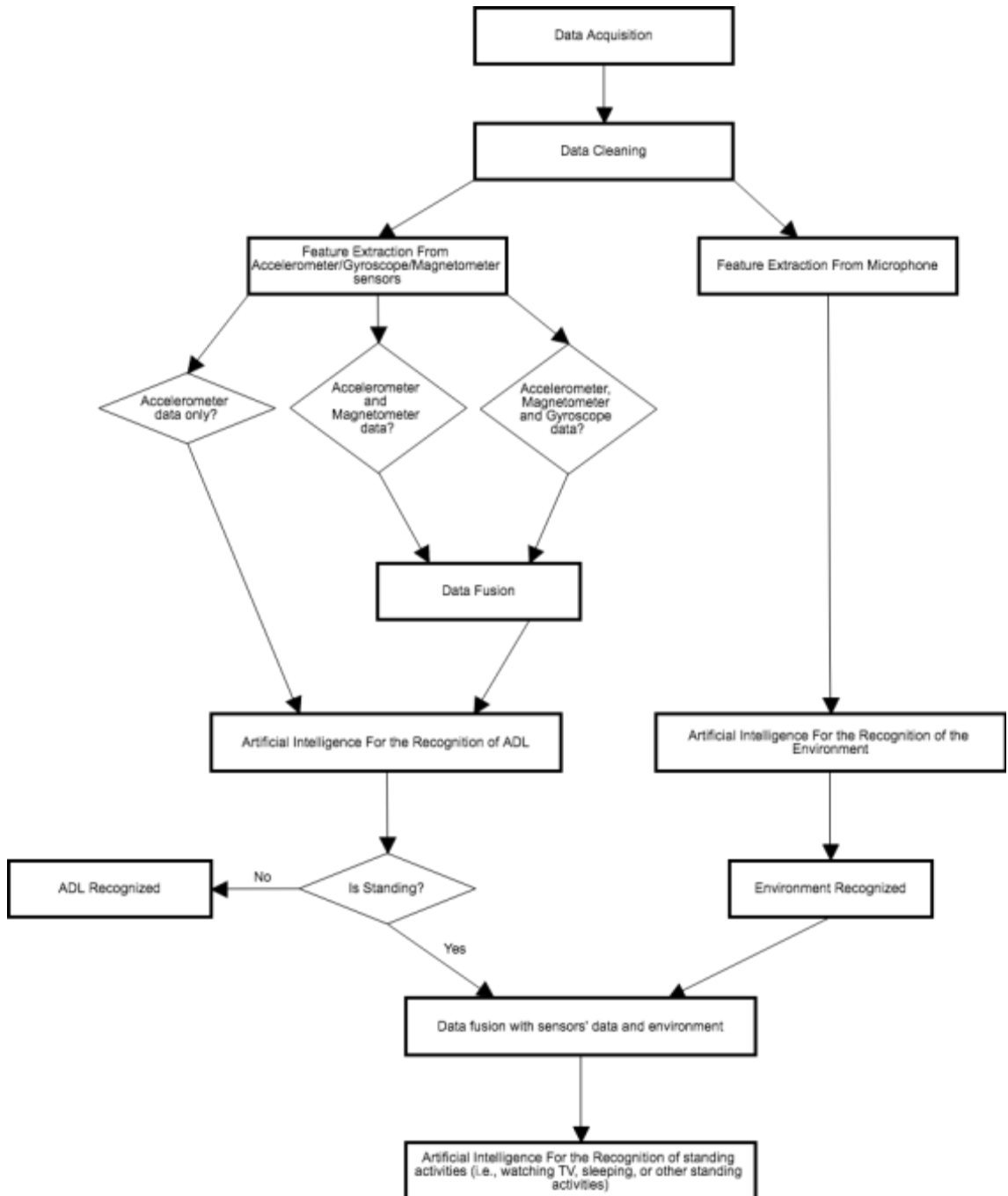

*Figure 14 - Simplified diagram for the framework for the identification of ADL.*

Firstly, based on the results obtained in the section 4.1, the best results achieved for each type of neural network are presented in the table 4, verifying that the best method for the recognition of the environments is the Feedforward neural networks with non-normalized data, reporting an accuracy of 86.50%.

Secondly, based on results obtained with the use of the environment recognized and the accelerometer data, presented in the section 4.2, the recognition of standing activities is allowed and the best results achieved for each type of neural network are presented in the table 5, verifying that the best method for the recognition of the standing activities is the DNN method with normalization of the data and the application of $L_2$ regularization, reporting an accuracy of 100%.

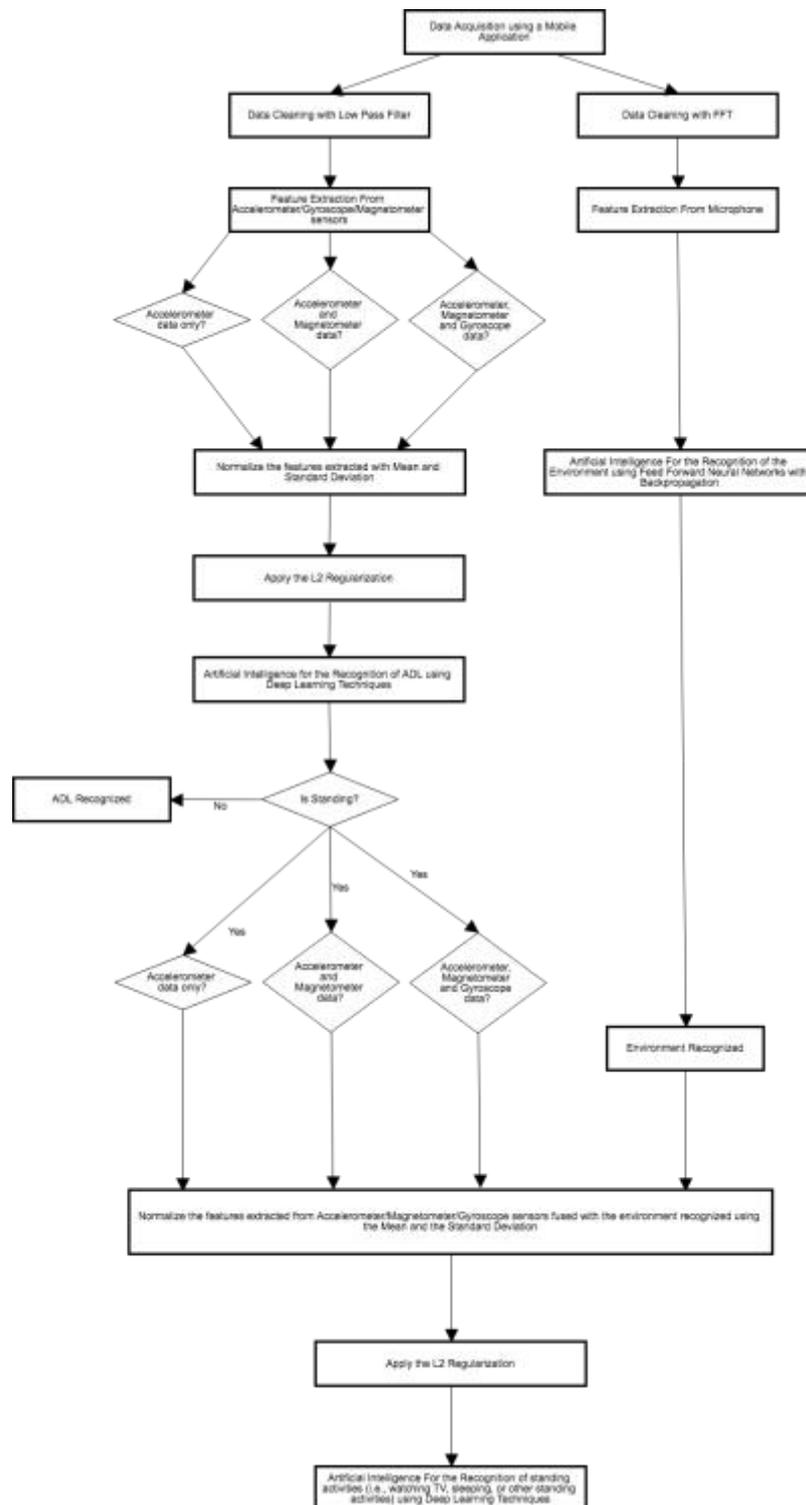

Figure 15 - Simplified diagram for the framework for the identification of ADL with indication of the methods for each stage.

Thirdly, based on results obtained with the use of the environment recognized and the accelerometer and magnetometer sensors' data, presented in the section 4.3, the recognition of standing activities is allowed and the best results achieved for each type of neural network are presented in the table 6, verifying that the best method for the recognition of the standing activities is the DNN method with normalization of the data and the application of $L_2$ regularization, reporting an accuracy of 100%.

Finally, based on results obtained with the use of the environment recognized and the accelerometer, magnetometer and gyroscope sensors' data, presented in the section 4.4, the recognition of standing activities is allowed and the best results achieved for each type of neural network are presented in the table 7, verifying that the best method for the recognition of the standing activities is the DNN method with normalization of the data and the application of $L_2$ regularization, reporting an accuracy of 100%.

In conclusion, when the activity was recognized as standing and the environment is correctly identified, the accuracy for the recognition of standing activities is 100%. As presented in the figure 15, at this stage of the development of the framework for the recognition of ADL and their environments, two different artificial intelligence methods are defined, these are:
- DNN with normalized data for the general identification of ADL;
- Feedforward neural networks with non-normalized data for the general identification of the environments;
- DNN with normalized data for the identification of standing activities.

## 6. Conclusions

The development of a framework for the recognition of ADL [1] and their environments using the sensors available in the off-the-shelf mobile devices, including accelerometer, gyroscope, magnetometer and microphone, with the architecture presented in [5-7] has several modules, such as data acquisition, data processing, data fusion and artificial intelligence methods. At this stage of the development, the proposed ADL for the recognition are running, walking, standing, going upstairs, sleeping, and going downstairs, and the proposed environments for the recognition are bar, classroom, gym, kitchen, library, street, hall, watching TV, and bedroom.

Depending on the types of sensors, several features were extracted from the sensors' data for further processing. The features extracted from the microphone are 26 MFCC coefficients, Standard Deviation of the raw signal, Average of the raw signal, Maximum value of the raw signal, Minimum value of the raw signal, Variance of the of the raw signal, and Median of the raw signal. And, the features extracted from the accelerometer, magnetometer and gyroscope sensors are the 5 greatest distances between the maximum peaks, the Average of the maximum peaks, the Standard Deviation of the maximum peaks, the Variance of the maximum peaks, the Median of the maximum peaks, the Standard Deviation of the raw signal, the Average of the raw signal, the Maximum value of the raw signal, the Minimum value of the raw signal, the Variance of the of the raw signal, and the Median of the raw signal. The method developed should be a function of the number of sensors available in the off-the-shelf mobile devices, and adapted to the limited resources of these devices.

In coherence with the previous studies [4, 14], this research includes the comparison of three different types of neural networks, such as MLP with Backpropagation using the Neuroph framework [15], the Feedforward Neural Network with Backpropagation using the Encog framework [16], and the DNN using DeepLearning4j framework [17], verifying that the DNN is the best method for the recognition of general ADL and standing activities, but the Feedforward Neural Network with Backpropagation is the best method for the recognition of environments.

The accuracies of the recognition ADL and their environments are different depending on the different stages of the framework for the recognition of ADL and environments. Firstly, the best accuracy for the recognition of the general ADL, presented in previous studies [4, 14], is 85.89%, implementing DNN using $L_2$ regularization and normalized data. Secondly, the best accuracy for the recognition of the environments is 86.50%, implementing Feedforward neural networks with Backpropagation using non-normalized data. Finally, the recognition of standing activities are always around 100% with all types of neural networks, but, due to the performance, the best method for the implementation in the framework is DNN using $L_2$ regularization and normalized data.

As future work, the methods for the recognition of ADL presented in this study should be implemented during the development of the framework for the identification of ADL and their environments, adapting the method to the number of sensors available on the mobile device. The recognition of the environments allows the framework for identify the location in the indoor/outdoor environments, where the ADL were performed. The recognition of the environment can also improve the recognition of ADL, increasing the number of ADL recognized. The data related to this research is available in a free repository [50].

**Acknowledgements**

This work was supported by FCT project **UID/EEA/50008/2013** (*Este trabalho foi suportado pelo projecto FCT UID/EEA/50008/2013*).

The authors would also like to acknowledge the contribution of the COST Action IC1303 – AAPELE – Architectures, Algorithms and Protocols for Enhanced Living Environments.